# A twist on folding: Predicting optimal sequences and optimal folds of simple protein models with the hidden-force algorithm

István Kolossváry[1, 2,*] and Kevin J. Bowers[3]

[1]*Department of Chemistry, Budapest University of Technology and Economics, H-1111 Budapest, Hungary*

[2]*BIOKOL Research, LLC, Madison, New Jersey 07940, USA*

[3]*Los Alamos National Laboratory, Los Alamos, New Mexico 87545, USA*

[*]Present address: D. E. Shaw Research, LLC, New York, NY 10036, USA; istvan@kolossvary.hu

We propose a new way of looking at global optimization of off-lattice protein models. We present a dual optimization concept of predicting optimal sequences as well as optimal folds. We validate the utility of the recently introduced hidden-force Monte Carlo optimization algorithm by finding significantly lower energy folds for minimalist protein models than previously reported. Further, we also find the protein sequence that yields the lowest energy fold amongst all sequences for a given chain length and residue mixture. In particular, for protein models with a binary sequence, we show that the sequence-optimized folds form more compact cores than the lowest energy folds of the historically fixed, Fibonacci-series sequences of chain lengths of 13, 21, 34, 55, and 89. We emphasize that while the protein model we used is minimalist, the methodology is applicable to detailed protein models, and sequence optimization may yield novel folds and aid de novo protein design.

I. INTRODUCTION

We recently introduced the hidden-force algorithm (HFA), a global Monte Carlo optimization method and used it to predict low-energy binary Lennard-Jones (BLJ) cluster configurations [1]. In this communication, we apply HFA to find the optimal fold of simple protein models. Further, we also find the protein sequence that yields the lowest energy fold amongst all sequences for a given chain length and residue mixture. In particular, we study protein AB (PAB) models [2, 3]. These have close similarity to BLJ models. Despite their minimalism, PAB models mimic a basic feature of protein folding—the formation of a hydrophobic core. Similar to BLJ models PAB models consist of only two types of residues, denoted A and B. The interaction energy between two A residues (AA interaction) is twice as strong as AB or BB interactions to promote core formation. The interaction potential has a pseudo Lennard-Jones form and the total potential energy includes angle-bending and torsion terms to account for local interactions. Historically, PAB models use a Fibonacci series sequence protein with chain lengths of 13, 21, 34, 55, and 89 [4, 5]. At first, we revisit the Fibonacci sequences and find that much lower energy folds exist than previously reported. Moreover, the new putative global minima show topological features qualitatively different from real proteins. The new low-energy folds are deeply knotted indicative of a flaw in the PAB model. Simplifying the local interaction terms introduced to extend the PAB model to three dimensions [3], we obtain new realistic low-energy folds without knots. Using this simplified potential we use HFA optimization to find non-Fibonacci sequences that fold into still lower energy structures with more compact cores comprised of A residues.

## II. MODEL AND ALGORITHMIC DETAILS

PAB models are one of the minimalist protein models [6]. We used the model by Irbäck *et. al* [3], a well-studied three-dimensional extension of the original two-dimensional PAB model by Stillinger *et. al* [2]. In this model proteins consist of two types of residues, A and B. Each residue represents a Cα atom. The Cα–Cα bonds are set to unit length and the potential energy is:

$$E = -\kappa_1 \sum_{i=1}^{N-2} \cos C\alpha_{i,i+1,i+2} - \kappa_2 \sum_{i=1}^{N-3} \cos C\alpha_{i,i+1,i+2,i+3} + \sum_{i=1}^{N-2} \sum_{j=i+2}^{N} 4\varepsilon(\sigma_i, \sigma_j) \left(\frac{1}{r_{ij}^{12}} - \frac{1}{r_{ij}^6}\right).$$

Eq. 1

The first two sums represent local interactions as angle-bending or torsion energies, involving three or four consecutive Cα atoms, respectively. The last double sum is a pseudo Lennard-Jones (LJ) potential representing long-range interactions. $N$ is the number of residues, $r_{ij}$ is the distance between two residues, $\varepsilon(\sigma_i, \sigma_j)$ is the residue pair-specific LJ well-minimum depth, and $\kappa_1$ and $\kappa_2$ are empirical parameters determined by Monte Carlo simulations to reproduce qualitatively Cα angle and torsion distributions of proteins in the Protein Data Bank (PDB) [3]. We used $\kappa_1 = -1$ and $\kappa_2 = 0.5$. $\varepsilon(\sigma_i, \sigma_j)$ favor the formation of a core of A residues analogous to the hydrophobic core of real proteins: $\varepsilon(A, A) = 1$ and $\varepsilon(A, B) = \varepsilon(B, B) = 0.5$. LJ interactions between adjacent residues are excluded. In lieu of constraints we add strong harmonic distance terms to keep Cα–Cα bond lengths at unity and flat-bottom angle-bending terms to disallow near linear Cα–Cα–Cα bond angles (within two degrees). Linear bond angles cause a catastrophic physical divergence in the torsion term of the PAB model.

The hidden-force algorithm [1] exploits that, though the gradient components of an additive potential sum to zero at a local minimum, each component's magnitude is generally nonzero. Disrupting this network of opposing forces (negative of the gradient) can result in the collective rearrangement of cluster atoms. Using a tug-of-war analogy to describe the basic HFA move, some players (atoms) simultaneously drop their ropes (drop their contributions to the potential). The remaining players then rearrange due to their net nonzero tugging and reach a partial impasse. Then the dropouts resume tugging until a new total impasse is achieved. There is no guarantee that the resulting cluster configuration will be lower in energy than the starting configuration, but we found HFA to be an exceptionally successful move set in a Monte Carlo cluster minimization. HFA trial configurations are highly dependent on the starting configurations since the moves are driven by forces already present—making the HFA Monte Carlo search non-Markovian.

Algorithmically, PAB models can be treated similar to BLJ clusters with slightly different LJ terms and additional geometric constraints favoring protein like chains. We used the same algorithm and software described in detail in [1] with two notable differences: (1) when the basic HFA move is applied to a given local minimum-energy configuration, only the pseudo LJ terms are dropped; the remaining terms stay in effect to preserve the chain geometry. (2) Single or multiple mutations are utilized by swapping the types of randomly selected residues rather than flipping the type of a single residue at a time, in order to keep the composition fixed while varying the sequence. A BLJ cluster of fixed size is fully determined by its A/B composition, but PAB models also depend on the *sequence* of the residues. Therefore, while swapping A and B particles in a BLJ cluster will not change its identity, swapping A and B residues in a PAB protein will be a mutation. Further, in [1], we found the optimal A/B composition of a BLJ cluster of fixed size with lowest energy. For PAB models, the optimal A/B composition is trivial, a sequence of all A residues (AA interactions are stronger than AB or BB interactions). The more interesting and challenging mutation study we carry out keeps the A/B composition fixed and optimizes over sequences to find the lowest energy fold.

### III. RESULTS AND DISCUSSION

To test HFA against existing optimization methods for PAB models, we run HFA searches on Fibonacci sequences in Table I and compare with putative global minimum-energy folds in [4, 5]. Table I clearly demonstrates that HFA Monte Carlo search is efficient searching the fold space of PAB models. The difference in energy does not carry information about structural differences. Direct comparison was, unfortunately, not possible because the coordinates of the structures reported in [4, 5] were not published. Nonetheless, based on the visualizations in [4, 5], FIG. 1 clearly shows that our new minima belong to a different topological class indicative of a flaw in PAB models. With the exception of S_13, which is simply too short and S_21 that forms a simple trefoil knot (one end of the chain folding back through a loop), every other fold is deeply knotted (coordinates are listed in the Supplementary Material). Knotted protein structures occur naturally [7, 8], however, this level of knot formation cannot be found in the PDB [9]. With $\kappa_1 = -1$ and $\kappa_2 = 0.5$ [3], the simple cosine terms favor a 180 degree bond angle and a zero degree torsion angle. In real proteins, the bond angle distribution has a well-defined structure and bond angles strictly fall within the range of 85-145 degrees [3]. Torsion angles, on the other hand, are more uniform; there are no disallowed values. Nevertheless, zero degree torsion angles are rare. Eq. 1 represents an additive potential and even though the PAB models will always be frustrated in three dimensions, we can expect that at or close to the global minimum many individual energy terms will be close to their minimum values. This is exactly what we can see in FIG. 1; numerous bond angles are close to 180 degrees (kept away from exact linearity by the flat bottom angle-bending term) and numerous torsion angles are very close to zero degrees. This is the geometrical basis for forming knots in the ground sate, never seen in real proteins.

Table I. Energies of the new putative global minima listed in ε units, found by HFA for five Fibonacci sequences. The * operator in the Sequence column means concatenation. The previously reported energies were taken from [4, 5].

| Model identifier with chain length | Sequence | Lowest energy [ε] previously reported | Putative global minimum energy [ε] |
|---|---|---|---|
| S_13 | ABBABBABABBAB | -26.507 | -27.171 |
| S_21 | BABABBAB * S_13 | -52.934 | -56.409 |
| S_34 | S_13 * S_21 | -98.357 | -106.115 |
| S_55 | S_21 * S_34 | -176.691 | -190.579 |
| S_89 | S_34 * S_55 | -311.614 | -325.578 |

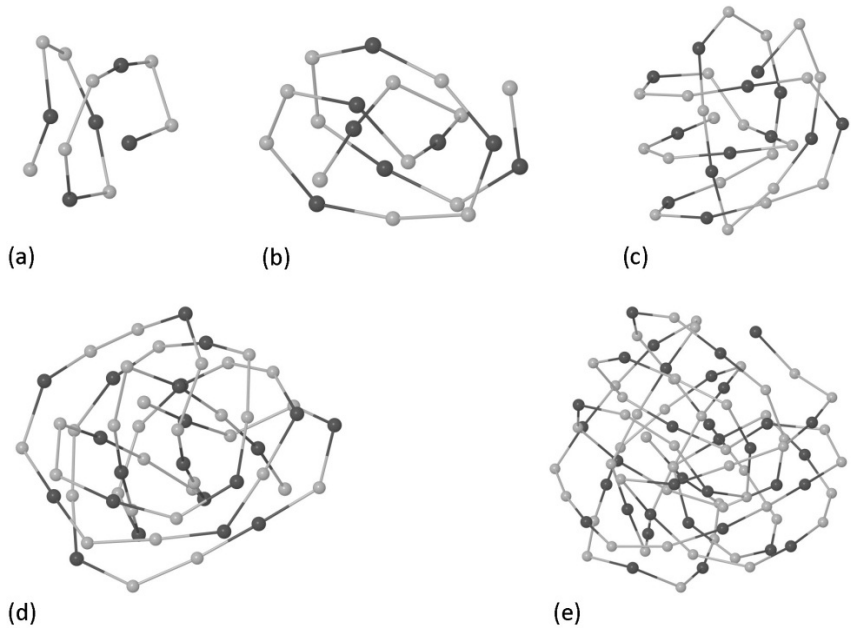

FIG. 1. Putative global minimum-energy folds found for 5 Fibonacci sequences using the potential in Eq. 1. Dark balls represent type A ("hydrophobic") residues and light grey balls are type B residues. Sequences and energies are given in Table I. (a) S_13, (b) S_21, (c) S_34, (d) S_55, and (e) S_89.

Ideally, more sophisticated potentials should be derived from the actual Cα bond angles and torsion angles found in the PDB via, e.g., the Boltzmann inversion method [6]. Keeping with the minimalist spirit of PAB models, however, we drop the angle-bending and torsion terms altogether, but we add a flat-bottom harmonic angle term that disallows Cα–Cα–Cα angles outside the 85-145 degree range. Using this potential we find that the Fibonacci sequences fold into globular structures with no tendency to form knots, and with a "hydrophobic" core. We then carried out sequence optimization using this potential. The left hand side of FIG. 2 shows the putative global minima found for the five Fibonacci sequences and the right hand side shows the lowest energy folds after sequence mutations were applied to the Fibonacci sequences (coordinates are available in the Supplementary Material). Visual inspection confirms the simplified PAB model yields compact folds with a clear core and no knots. The cores are more compact in the mutated sequences, which is quantified in Table II by the radius of gyration of the core (type A) residues. Table II also lists the energy drop after sequence optimization with respect to the putative global minimum energy of the Fibonacci sequence. We emphasize that while this potential is minimal, the methodology is applicable to detailed protein models, and sequence optimization may aid de novo protein design.

Table II. Fold optimization via sequence mutation. Energy drop and radius of gyration of the "hydrophobic" core formed by the type A residues. Energy is computed with the LJ-only potential (see text). The optimal sequences, coordinates, and absolute energies are listed in the Supplementary Material.

| Model identifier with chain length | Energy drop [ε] relative to Fibonacci sequence global min. | Radius of gyration of the core in Fibonacci sequence global min. | Radius of gyration of the core in optimal-sequence global min. |
|---|---|---|---|
| S_13 | -0.346 | 0.203 | 0.156 |
| S_21 | -1.369 | 0.288 | 0.223 |
| S_34 | -2.748 | 0.336 | 0.281 |
| S_55 | -9.074 | 0.371 | 0.327 |
| S_89 | -18.576 | 0.460 | 0.365 |

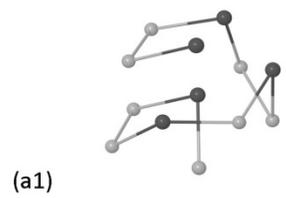
(a1)
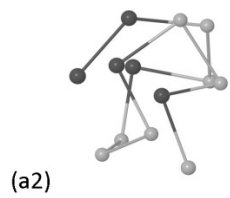
(a2)

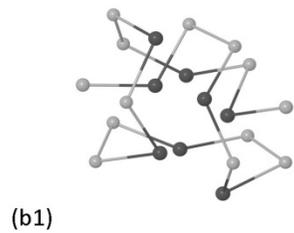
(b1)
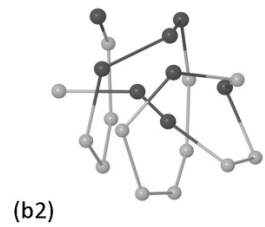
(b2)

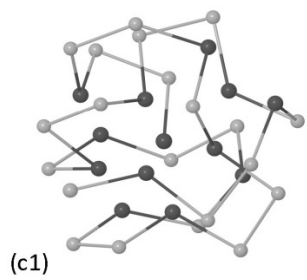
(c1)
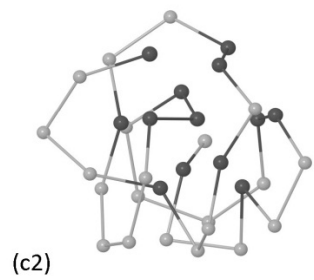
(c2)

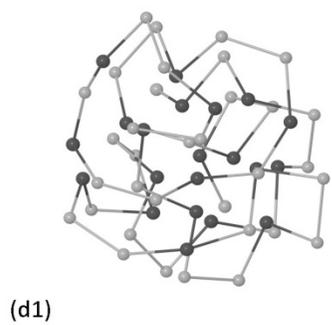
(d1)
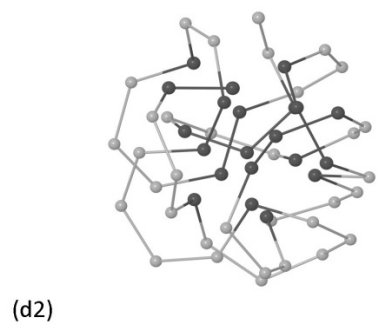
(d2)

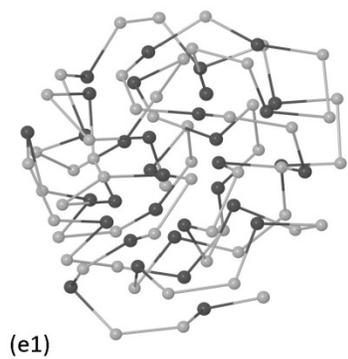
(e1)
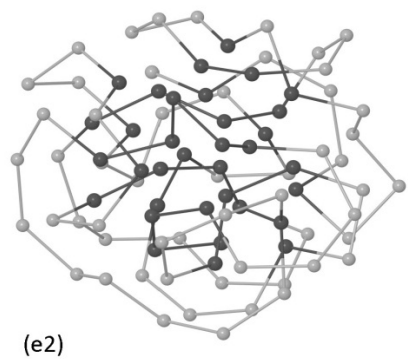
(e2)

FIG. 2. Putative global minimum-energy folds using the simplified LJ-only potential (see text). On the left hand side the Fibonacci sequences are shown and on the right hand side the optimal folds are displayed that were found after sequence optimization. Dark balls represent type A ("hydrophobic") residues and light grey balls are type B residues. The optimal sequences, coordinates, and energies are listed in the Supplementary Material. (a1, a2) S_13, (b1, b2) S_21, (c1, c2) S_34, (d1, d2) S_55, and (e1, e2) S_89.

## IV. SUMMARY AND PERSPECTIVE

In this communication, we propose a dual concept for the global optimization of off-lattice protein models predicting optimal sequences as well as optimal folds. Using hidden-force Monte Carlo optimization we find the optimal fold of simple protein models. Further, we also find the protein sequence that yields the lowest energy fold amongst all sequences for a given chain length and residue mixture. In particular, we find that with a simplified potential, binary protein AB models fold into globular, compact structures with no tendency to form knots, and with a "hydrophobic" core. Historically, these models use a Fibonacci series sequence protein with chain lengths of 13, 21, 34, 55, and 89. We show that sequence optimization yields non-Fibonacci sequences that fold into still lower energy structures with more compact cores than the original Fibonacci sequences. We emphasize that while this is a minimalist protein model, the methodology is applicable to detailed protein models, and sequence optimization may yield novel folds and aid de novo protein design. Sequence optimization also hints at a hypothesis regarding the evolution of proteins. The number of different proteins manifest in nature is minuscule in view of the total number of potential sequences. The success of sequence optimization suggests that sequences affording stable folds can be pre-selected solely by minimizing their free energy. Protein sequences for specific protein functions could, then, evolve in finite time probing all the stable folds, rather than all possible sequences. Our current research is focusing on multi-scale protein modeling and molecular dynamics simulations to bring sequence optimization to the all atom level.

# Supplementary Material

Description:

List of coordinates, sequences, and energies of putative global minimum folds of PAB models. The coordinates are Cartesian coordinates. The sequence is identified by the residue type A or B in the first column. Energies are absolute energies in epsilon units. The first 5 models were computed with the potential in Eq. 1. The other 5 models were computed with the simplified potential including the pseudo Lennard-Jones potential, the bond length restraint and the flat-bottom angle bending constraint to keep the Calpha-Calpha-Calpha bond angles within the 85-145 degree range.

```
S_13 Fibonacci sequence global minimum energy computed by Eq. 1 =  -27.171382
A     -2.210729   -0.766557    0.859473
B     -1.468936   -0.515469    1.481321
B     -1.125457    0.413991    1.346684
A     -2.015611    0.307761    0.903583
B     -2.875024    0.042923    0.466239
B     -2.923165   -0.824726   -0.028602
A     -2.104003   -1.398285   -0.026841
B     -1.292453   -1.270099    0.543208
A     -1.100612   -0.289674    0.498904
B     -1.256479    0.656395    0.214902
B     -2.203998    0.733500   -0.095357
A     -1.998733   -0.244823   -0.122725
B     -1.256899   -0.823838   -0.460991

S_21 Fibonacci sequence global minimum energy computed by Eq. 1 =  -56.408990
B     -3.502088   -2.347106   -2.580986
A     -3.633583   -3.070794   -1.903502
B     -3.182330   -2.964804   -1.017422
A     -2.558379   -2.207111   -0.826145
B     -1.942815   -1.419028   -0.828706
B     -1.631517   -1.022569   -1.692370
A     -1.917468   -1.469613   -2.539947
B     -2.396659   -2.287089   -2.859494
A     -2.809866   -3.188090   -2.727356
B     -2.658749   -3.616882   -1.836679
B     -2.088394   -3.137323   -1.169805
A     -1.497975   -2.495615   -0.680298
B     -0.926488   -1.760655   -1.045314
B     -0.814412   -1.662030   -2.034107
A     -1.235498   -2.340732   -2.635811
B     -1.870827   -3.002000   -2.236963
A     -2.685908   -2.555420   -1.867897
B     -3.446519   -2.008019   -1.518868
B     -2.589028   -1.522775   -1.689879
A     -1.763180   -2.085958   -1.661572
B     -0.995117   -2.724986   -1.620079

S_34 Fibonacci sequence global minimum energy computed by Eq. 1 = -106.115214
A     -0.054962    2.229487   -0.900343
B     -0.853011    2.143012   -1.496697
```

```
B    -0.951238    1.732429   -0.590181
A    -0.524775    1.446744    0.268022
B     0.330605    1.385370    0.782374
B     1.265892    1.429216    1.133538
A     1.199972    1.905059    0.256483
B     0.995707    2.354906   -0.612946
A     0.707322    2.800658   -1.460375
B    -0.237086    3.129292   -1.450697
B    -0.878686    2.917563   -0.713459
A    -0.848802    2.520503    0.203846
B    -0.411806    2.204694    1.046045
B     0.565968    2.364524    1.181737
A     1.553768    2.506594    1.117971
B     2.284360    2.032034    0.627028
A     2.006677    1.145438    0.257113
B     1.079417    0.771783    0.233288
B     0.127003    0.470460    0.187356
A    -0.495836    0.717894   -0.554834
B    -0.200073    1.245807   -1.350967
A     0.708342    1.655607   -1.433690
B     1.614781    2.075060   -1.482927
B     1.824123    2.954070   -1.054539
A     1.063933    3.448186   -0.632687
B     0.147975    3.147931   -0.366482
B     0.176094    2.278401    0.126597
A     0.388799    1.417582   -0.335729
B     0.682145    0.618698   -0.860826
A     1.419258    1.273341   -0.693175
B     2.059386    2.006824   -0.464622
B     1.809237    2.833024    0.040171
A     0.893430    3.031008    0.389595
B    -0.039220    3.183399    0.716612

S_55 Fibonacci sequence global minimum energy computed by Eq. 1 = -190.579174
B    -1.048644   -1.434770   -2.612892
A    -0.857230   -1.125885   -1.681255
B    -0.596707   -0.675295   -0.827388
A    -0.290393   -0.155407   -0.029967
B     0.110010    0.484079    0.626335
B     0.162770    0.776914   -0.328370
A     0.162639    0.678281   -1.323493
B     0.219978    0.432508   -2.291121
A     0.444364    0.055716   -3.189829
B     1.268189   -0.503306   -3.283678
B     1.459310   -1.259428   -2.657776
A     0.965495   -1.656927   -1.884379
B     0.255341   -1.879653   -1.216491
B    -0.653161   -1.783533   -0.809818
A    -1.504582   -1.262869   -0.746650
B    -1.578376   -0.369291   -1.189448
A    -1.557596    0.523631   -1.639172
B    -0.739898    1.066675   -1.830141
B     0.113194    1.559915   -2.000266
A     1.027615    1.166839   -1.903716
B     1.928092    0.747291   -1.789189
A     2.048504    0.042246   -1.090328
B     1.410687   -0.255968   -0.380218
B     0.694872   -0.476615    0.282294
A    -0.074711   -0.632197    0.901598
B    -0.858279   -0.013484    0.844884
B    -0.934677    0.677467    0.126029
A    -0.778999    0.418222   -0.827153
B    -0.560743   -0.043843   -1.686720
```

```
A       -0.310408   -0.533405   -2.521979
B        0.337372   -1.040818   -3.090234
B        0.770256   -0.531522   -2.346440
A        1.031028    0.081018   -1.600259
B        1.181685    0.761207   -0.882872
B        0.607766    1.578979   -0.926071
A       -0.391078    1.531446   -0.918889
B       -1.367073    1.318376   -0.873801
A       -1.829006    0.501407   -0.528579
B       -1.397912   -0.309658   -0.133193
B       -0.837781   -1.075151    0.183464
A        0.074510   -1.212677   -0.202294
B        0.924085   -1.176072   -0.728487
A        1.540988   -0.860680   -1.449567
B        1.854634   -0.317133   -2.228143
B        1.320831    0.424115   -2.635087
A        0.699667    1.130165   -2.975156
B       -0.300051    1.108245   -2.966151
B       -0.904839    0.382850   -2.637471
A       -1.426102   -0.387277   -2.269794
B       -1.917393   -1.171891   -1.891624
A       -1.532661   -2.027320   -1.544895
B       -0.607225   -2.241704   -1.857312
B       -0.015642   -1.517680   -2.212015
A        0.222705   -0.833229   -1.523018
B        0.419780   -0.197002   -0.777114

S_89 Fibonacci sequence global minimum energy computed by Eq. 1 = -325.578412
A       -0.890676    5.222297   11.293380
B       -1.668704    4.931778   10.736363
B       -2.158498    5.528157   10.100414
A       -1.668164    6.004270    9.370424
B       -1.051123    6.389643    8.684315
B       -0.152088    6.170183    8.305408
A        0.750047    5.925282    7.950197
B        1.538695    5.358181    8.187744
A        0.826827    4.723130    7.887825
B        0.036563    4.606234    7.286312
B       -0.894691    4.884079    7.050588
A       -1.682813    4.670039    7.627694
B       -2.347825    4.340134    8.297708
B       -2.648735    3.820308    9.097229
A       -2.285703    3.421407    9.939301
B       -1.657824    3.205639   10.687106
A       -1.047698    2.438666   10.488372
B       -0.448592    1.679773   10.233127
B       -0.073118    1.635437    9.307356
A        0.175184    2.320731    8.622727
B        0.429502    3.028740    7.963908
A        0.961370    3.773140    8.367617
B        1.461699    4.496864    8.842896
B        1.060899    4.700943    9.736040
A        0.580722    4.800837   10.607502
B        0.051615    4.760676   11.455104
B       -0.206286    3.937065   11.960229
A        0.307527    3.094937   11.796475
B        0.935780    2.921734   11.037992
A        1.503541    2.870146   10.216418
B        2.004594    2.923607    9.352655
B        1.989546    3.553174    8.575855
A        1.817329    4.208172    7.840111
B        1.556098    4.869972    7.137416
B        0.721844    5.407230    7.013423
```

```
A     -0.183812    5.670363    7.345912
B     -0.999588    5.548707    7.911342
A     -1.748901    5.296576    8.523683
B     -2.477520    5.020295    9.150407
B     -2.337992    4.467211    9.971765
A     -1.423481    4.092965   10.125423
B     -0.656610    4.211696    9.494699
A      0.233414    4.007857    9.902503
B      1.076682    3.845623   10.414925
B      1.784226    4.498139   10.686219
A      1.493482    5.421480   10.435407
B      0.703605    5.818557    9.968050
B     -0.134302    6.070308    9.483763
A     -0.743820    5.424754    9.023605
B     -0.914901    4.602955    8.480119
A     -0.076806    5.105053    8.266837
B      0.622976    5.476910    8.876777
B      1.522606    5.563877    9.304680
A      2.156608    4.894567    9.692060
B      1.890921    3.930661    9.709134
A      1.058479    3.482964    9.382628
B      0.138880    3.359732    9.009600
B     -0.812989    3.507476    8.741054
A     -1.610013    4.038142    9.029405
B     -1.492645    4.909885    9.505099
B     -0.979113    5.537814   10.089894
A     -0.136436    5.670471   10.611711
B      0.713138    5.769108   11.129872
A      1.235614    4.994903   11.487115
B      0.879173    4.060589   11.489658
B      0.131152    3.711345   10.925309
A     -0.587034    3.420047   10.293368
B     -1.290857    3.148875    9.636790
B     -1.913205    2.998510    8.868630
A     -1.660815    3.509169    8.046727
B     -0.894610    3.973606    7.602627
A     -0.054379    4.003275    8.144040
B      0.339086    4.438247    8.953967
B      0.011977    5.023939    9.695563
A     -0.501837    4.623321   10.454180
B     -0.959711    4.067838   11.148288
A     -0.714692    3.117182   11.338593
B     -0.066216    2.493851   10.901628
B      0.583538    2.018076   10.308790
A      0.927769    2.356424    9.432990
B      1.249645    2.721486    8.559421
B      1.494238    3.149799    7.689523
A      0.878466    3.814968    7.267178
B     -0.065908    3.506738    7.152503
A     -0.606076    2.976902    7.806334
B     -0.954466    2.474637    8.597758
B     -1.161863    2.043503    9.475885
A     -0.346425    2.607860    9.604576
B      0.415302    3.024958   10.100354

S_13 Fibonacci sequence global minimum energy computed by LJ term =   -18.264451
A     -1.872515   -0.893300    1.154829
B     -2.731132   -0.568108    0.758563
B     -2.520868    0.363275    1.055738
A     -1.590969    0.182426    1.376019
B     -1.309534    0.775524    0.621679
B     -0.785642    0.122565    0.074705
A     -0.850686   -0.469205    0.878182
```

```
B    -1.174009   -1.161829    0.233409
A    -2.165186   -1.224768    0.116772
B    -2.796179   -0.659337   -0.414391
B    -2.608159    0.249817   -0.042793
A    -1.790042   -0.174069    0.345796
B    -1.637280   -0.521693   -0.579308

S_21 Fibonacci sequence global minimum energy computed by LJ term =   -38.280724
B    -2.598970   -1.071744   -0.989501
A    -2.555464   -1.832540   -1.637030
B    -3.179306   -2.381891   -1.081124
A    -3.006899   -3.100627   -1.754692
B    -2.258990   -3.762954   -1.710463
B    -1.448311   -3.577823   -1.155012
A    -1.257632   -2.619644   -0.941625
B    -0.956978   -2.301922   -1.840880
A    -1.693473   -1.901240   -2.385884
B    -1.498008   -2.638748   -3.032317
B    -2.233398   -3.230214   -2.701607
A    -2.734630   -2.377477   -2.554629
B    -3.551075   -2.074591   -2.063023
B    -3.307601   -1.117647   -1.904976
A    -2.600028   -1.274972   -2.593879
B    -1.969334   -0.931424   -1.898036
A    -1.564488   -1.608399   -1.283378
B    -2.179463   -2.041624   -0.624500
B    -2.319442   -3.001169   -0.868793
A    -1.967465   -2.733347   -1.765666
B    -1.265945   -3.308704   -2.186183

S_34 Fibonacci sequence global minimum energy computed by LJ term =   -75.170999
A     0.472359    2.418285   -0.638095
B    -0.322102    2.299673   -1.233713
B    -0.047300    1.650925   -1.943368
A     0.691744    1.410112   -1.314225
B     0.478766    0.496016   -0.969161
B    -0.167306    0.439314   -0.207994
A    -0.553580    1.238591    0.252382
B     0.290259    1.228340    0.788878
A     0.371417    2.187270    0.517094
B     0.329232    2.937186    1.177280
B     1.087189    3.424014    0.743116
A     1.236341    2.671553    0.101590
B     2.127574    2.239059    0.238152
B     2.271682    2.431017   -0.732612
A     1.412721    2.888905   -0.961792
B     0.950592    3.588052   -0.416237
A     0.206155    3.177982    0.110693
B    -0.407799    3.236765   -0.676456
B     0.371919    3.267500   -1.301832
A     0.730286    2.427743   -1.709729
B     1.608837    1.981712   -1.538835
A     1.389475    1.818740   -0.576900
B     1.491686    0.851222   -0.808113
B     0.968414    0.654887    0.021127
A     0.390090    1.417616   -0.268326
B    -0.318322    1.226465   -0.947748
B    -1.130292    1.717089   -0.631537
A    -0.481968    2.287088   -0.126787
B    -0.645757    2.798181    0.716986
A    -0.449328    1.931999    1.176489
B     0.464218    1.940154    1.583146
B     1.267028    2.344961    1.145395
```

```
A        1.267005    1.595162    0.483729
B        2.057765    1.152922    0.060503

S_55 Fibonacci sequence global minimum energy computed by LJ term = -137.241642
B        1.373813    0.169900   -0.003080
A        0.839574    0.156997   -0.848314
B        1.224370    0.867832   -1.437077
A        0.377357    0.770842   -1.959722
B       -0.189950    1.593290   -2.001428
B       -0.752958    1.436535   -1.189978
A       -0.408800    1.311824   -0.259385
B       -1.308185    0.916183   -0.073446
A       -0.858252    0.037826   -0.234832
B       -1.284447   -0.789417    0.131262
B       -0.695297   -1.273802   -0.515477
A       -0.006945   -0.549511   -0.555161
B        0.858993   -0.783863   -0.113315
B        0.673085   -0.297359    0.740354
A       -0.288062   -0.478856    0.532380
B        0.025050   -1.402222    0.310224
A        0.334731   -1.623659   -0.614471
B       -0.271404   -2.223190   -1.137120
B       -1.125174   -1.780380   -1.410972
A       -1.487109   -1.038236   -1.975084
B       -1.530026   -0.039385   -1.996363
A       -0.665714    0.460180   -1.938074
B       -1.205005    1.252159   -2.224309
B       -1.998532    0.927871   -1.709381
A       -1.323163    0.539623   -1.082374
B       -1.904416   -0.024012   -0.495467
B       -1.661492   -0.902707   -0.906419
A       -0.844182   -0.435675   -1.243885
B       -0.228644   -1.185260   -1.487269
A        0.415769   -0.978629   -2.223498
B        0.308049   -1.404688   -3.121756
B       -0.235283   -0.595015   -3.343612
A        0.259806    0.006437   -2.716602
B        1.061168   -0.536598   -2.967464
B        1.955183   -0.345303   -2.562321
A        1.681263   -1.004307   -1.861835
B        1.202204   -1.648374   -2.458224
A        0.675894   -1.879328   -1.639899
B       -0.125260   -1.963959   -2.232342
B       -0.771985   -1.419525   -2.766510
A       -0.580275   -0.552099   -2.307364
B       -0.939675    0.136461   -2.937217
A       -0.345111    0.937011   -2.862293
B        0.600021    1.261688   -2.898470
B        1.277340    0.561780   -2.671830
A        1.081183   -0.097319   -1.945809
B        1.974153    0.069299   -1.527669
B        1.666049   -0.570189   -0.823307
A        0.795558   -0.858027   -1.222548
B        0.130805   -0.203018   -1.581805
A       -0.201357    0.515911   -0.971229
B        0.326022    1.360430   -1.064280
B        0.837993    1.095576   -0.247129
A        0.201303    0.373024    0.022210
B       -0.534534    0.612033    0.655785

S_89 Fibonacci sequence global minimum energy computed by LJ term = -242.825509
A        0.262401    4.242779    8.095651
B        1.068432    4.121103    7.516420
```

```
B     0.659709    3.332733    7.056616
A    -0.176719    3.349537    7.604434
B    -1.092642    3.300990    7.206027
B    -0.785156    4.044326    6.611969
A     0.099885    4.260447    7.024268
B     0.580675    5.096724    7.287862
A     1.005547    5.043997    8.191578
B     1.610226    4.300315    8.476709
B     1.707206    3.305931    8.519077
A     0.790503    3.287169    8.119952
B     0.917618    2.416369    7.645033
B    -0.020586    2.336864    7.308208
A    -0.820510    2.488857    7.888742
B    -1.047035    3.375591    8.291715
A    -0.135793    3.408852    8.702235
B     0.110602    2.490435    8.392721
B     0.127264    1.999318    9.263654
A    -0.024691    2.898665    9.673627
B    -0.906480    3.115770   10.092323
A    -1.040584    3.705501    9.295941
B    -1.951092    3.357325    9.072896
B    -1.708706    2.545623    8.541497
A    -0.829295    2.606496    9.013652
B    -0.838998    2.075203    9.860784
B    -0.099781    2.182613   10.525629
A    -0.039922    3.167379   10.688878
B    -0.649078    3.759855   11.216037
A    -0.886562    4.157247   10.329651
B    -1.805176    3.938417   10.000617
B    -2.328988    4.370486    9.266494
A    -1.628394    4.257806    8.561887
B    -1.998209    3.625284    7.881336
B    -1.661353    4.286907    7.211425
A    -0.799669    4.226626    7.715275
B    -0.551235    5.086298    7.268911
A    -0.073144    5.290318    8.123196
B     0.571077    6.037373    7.959217
B     0.517272    6.824111    8.574157
A     0.401977    6.758480    9.565318
B     1.204549    6.162898    9.531305
A     0.617416    5.492169    9.984507
B     0.713704    6.080373   10.787467
B    -0.234202    6.389185   10.865619
A    -0.345923    5.994866    9.953464
B    -0.724722    6.652212    9.302000
B    -0.402554    6.266276    8.437558
A     0.307962    5.836758    8.994945
B     0.333013    4.837120    9.004377
A    -0.571119    4.495350    8.747993
B    -1.120092    5.123399    8.196463
B    -1.849419    5.310010    8.854684
A    -1.339882    4.756065    9.513103
B    -2.071029    4.998725   10.150706
A    -1.718019    4.498096   10.941120
B    -0.872181    4.798858   11.381687
B     0.077675    4.558385   11.581548
A     0.182024    4.230095   10.642753
B     1.108881    3.982703   10.925125
B     1.911000    4.295324   10.416331
A     1.632244    5.247700   10.292738
B     0.902394    5.026502   10.939569
A    -0.052887    5.286314   10.798382
B    -0.303312    5.670244   11.687135
```

```
B    -1.181011    5.874377   11.253575
A    -1.080298    5.286839   10.450671
B    -1.417309    5.856847    9.701329
B    -0.703314    5.492788    9.103273
A    -0.296294    4.909130    9.805896
B    -0.038124    3.964528    9.603241
A     0.775288    3.919465    9.023301
B     1.664446    3.811010    9.467861
B     2.195119    4.657098    9.417666
A     1.314974    5.095290    9.235099
B     0.926354    4.483003    9.923632
A     0.806105    3.497458   10.042964
B     1.700779    3.211873   10.386473
B     1.777215    2.716025    9.521436
A     0.879318    2.857280    9.104515
B     1.042755    1.975180    8.662719
B     1.157882    1.628993    9.593793
A     0.871987    2.423081   10.130158
B     0.909424    2.861616   11.028093
A     0.419355    3.524880   11.593702
B    -0.266002    2.803327   11.691918
B    -0.983369    2.604411   11.024224
A    -1.597429    3.379182   10.873698
B    -1.932869    2.849475   10.094665

S_13 Optimized sequence global minimum energy computed by LJ term =   -18.610603
A     1.038830   -0.765684    0.479799
A     1.015785   -0.076582   -0.244499
B     0.669287   -0.248282   -1.166702
B    -0.056089   -0.909581   -0.975615
A     0.055459   -0.625012   -0.023471
B     0.255497   -1.586615    0.164427
B     1.221177   -1.841911    0.212231
B     1.814616   -1.696593   -0.579420
A     1.903484   -0.763329   -0.231396
B     1.786908   -0.246508   -1.079514
B     1.376632   -1.087295   -1.432712
A     0.902306   -1.121712   -0.553038
B     0.741207   -2.082997   -0.776590

S_21 Optimized sequence global minimum energy computed by LJ term =   -39.650163
B    -1.168606   -0.122705   -2.808934
A    -1.862785    0.435824   -2.354887
A    -2.028844    0.286177   -1.380195
B    -2.208717    0.787195   -0.533660
B    -2.321778    1.581406   -1.130690
A    -2.688791    1.008135   -1.863261
B    -3.071911    0.417466   -1.153099
A    -2.710029   -0.513141   -1.098203
B    -1.740871   -0.628266   -0.880305
B    -1.228007    0.201424   -0.659886
B    -1.321287    1.055575   -1.171476
B    -1.642009    1.469229   -2.023548
B    -2.236188    1.350234   -2.819030
A    -2.846012    0.559679   -2.875036
A    -2.806989   -0.083118   -2.109995
A    -1.972821   -0.621171   -1.988909
B    -1.118503   -0.184810   -1.706544
B    -0.832964    0.693723   -2.089489
B    -1.179475    1.016090   -2.970403
B    -1.880494    0.483777   -3.444972
A    -2.235819   -0.357323   -3.037179
```

```
S_34 Optimized sequence global minimum energy computed by LJ term =  -77.919140
A     -1.545522    1.177180    0.136790
B     -1.852461    0.277788   -0.174471
B     -2.305021   -0.428425    0.370003
B     -3.101351    0.175172    0.330931
A     -3.195326    0.957649    0.946476
B     -4.038946    1.291080    1.367341
B     -4.224617    1.700988    0.474313
A     -3.255779    1.948018    0.456213
B     -2.867313    2.642656   -0.149237
A     -2.003360    2.160413   -0.004241
A     -1.353174    2.093840    0.752612
B     -0.702807    1.334230    0.756818
B     -1.148185    0.439531    0.790805
A     -2.144132    0.521769    0.827252
B     -2.763715    0.018408    1.429532
B     -3.745329    0.034185    1.239312
B     -4.162410    0.586755    0.517710
B     -3.599110    1.065813   -0.155487
A     -2.615162    1.147361    0.003235
A     -2.300719    1.565377    0.855517
A     -1.531879    1.230223    1.400088
B     -1.698313    0.324516    1.789952
B     -2.423976    0.230089    2.471492
B     -2.808893    1.058299    2.878807
B     -2.591776    1.960076    2.505099
A     -2.020424    2.161472    1.709489
A     -2.436276    2.622496    0.925573
B     -3.342021    3.018557    0.774705
B     -3.946384    2.397500    1.273737
A     -3.103517    1.925973    1.533034
B     -3.622960    1.621311    2.331381
B     -3.408948    0.689301    2.038883
A     -2.513309    1.063787    1.798910
B     -1.783385    1.210412    2.466526

S_55 Optimized sequence global minimum energy computed by LJ term = -146.316053
A     -2.973835    0.271336   -0.744174
A     -2.379573   -0.505505   -0.535926
B     -2.754117   -1.430975   -0.592679
B     -3.648686   -1.649556   -0.202848
B     -3.695617   -1.228583    0.703010
A     -3.529803   -0.268431    0.927986
B     -3.858544    0.200964    1.747496
B     -4.795079    0.534182    1.856442
B     -5.226262    1.098880    1.152737
B     -5.657353    0.725177    0.331451
B     -5.332960   -0.127324   -0.078441
A     -4.412782   -0.344033    0.247606
B     -4.467569   -0.661575    1.194267
B     -3.727868   -0.898194    1.824230
B     -2.825390   -0.967957    1.399183
B     -2.667484   -0.976216    0.411764
A     -3.408784   -0.604395   -0.147004
A     -3.362471   -0.704323   -1.140919
B     -2.573241   -0.313214   -1.614362
B     -1.827804    0.182300   -1.168505
A     -2.052405    0.686429   -0.334595
B     -1.634234   -0.087315    0.141275
B     -1.988619   -0.352474    1.037991
B     -2.719519    0.142977    1.507366
A     -3.266001    0.829831    1.028211
A     -2.887597    1.186242    0.173939
```

```
B    -3.195715    2.069915   -0.178455
B    -3.631644    2.059588   -1.078376
B    -3.574188    1.270052   -1.689385
A    -3.540737    0.270685   -1.677332
A    -4.301702   -0.377270   -1.710315
A    -5.072404   -0.159082   -1.111640
B    -5.432354    0.717969   -0.793489
A    -4.650887    0.712986   -0.169563
B    -5.035067    1.568016    0.178760
B    -4.598191    2.046878    0.940226
B    -4.142720    1.313925    1.445523
A    -4.369110    0.524691    0.874686
B    -5.225473    0.021948    0.992555
B    -5.377799   -0.910216    0.664122
B    -4.656841   -1.377197    0.152117
A    -4.358333   -0.925402   -0.688582
A    -4.064674    0.028497   -0.750527
A    -4.475313    0.734922   -1.327012
B    -4.626157    1.593895   -0.837716
B    -4.254866    2.224542   -0.156227
A    -3.969040    1.409887    0.348391
B    -3.388986    1.976633    0.933481
B    -2.482623    1.596761    1.118423
B    -2.082081    0.738179    0.798419
A    -2.703822    0.106200    0.335778
A    -3.627577    0.416598    0.111441
A    -3.686799    1.075113   -0.638791
B    -2.761678    1.354487   -0.895892
B    -2.574822    0.783251   -1.695125

S_89 Optimized sequence global minimum energy computed by LJ term = -261.401077
A     5.335687    2.905267   -9.716160
A     5.832940    3.467213   -9.055139
A     6.573647    2.963723   -8.610340
A     6.588762    1.996723   -8.356021
B     5.847304    1.652888   -7.779812
B     5.450839    0.761876   -8.000976
B     5.614742    0.326729   -8.886291
B     6.455216    0.556758   -9.376892
A     6.901018    1.437602   -9.217604
A     7.452963    1.296742   -8.395709
A     8.130765    1.874034   -8.851024
A     7.343220    2.415189   -9.145851
A     7.573687    3.296985   -8.734368
B     8.399980    3.828942   -8.919462
B     7.784366    4.611600   -9.011471
B     6.863774    4.955809   -8.826990
B     5.981103    4.509978   -8.678237
B     6.107616    3.756833   -8.032664
B     5.201538    3.341739   -8.114643
B     4.914555    2.436884   -7.800192
B     4.961990    1.641023   -8.403811
A     5.822732    1.383511   -8.842915
B     6.484006    0.839015   -8.326934
B     6.836661    1.268568   -7.495599
B     6.784960    2.248266   -7.301905
B     6.008944    2.723401   -7.716685
A     5.606019    2.472408   -8.596829
A     6.184722    2.264265   -9.385352
A     6.065528    1.427481   -9.919749
B     5.418240    0.665944   -9.952566
B     4.790851    0.986499   -9.242901
A     5.132237    1.897241   -9.475287
```

```
B    4.599280    2.534158    -8.918247
B    4.710900    3.517102    -9.064396
B    5.232536    4.180442    -9.600936
A    5.942526    3.752122   -10.159911
A    6.116817    3.564621   -11.126589
B    5.343402    3.660016   -11.753270
B    5.037800    2.712679   -11.848961
A    5.925344    2.478301   -11.452313
B    6.425150    3.088922   -12.066588
B    6.376102    4.087309   -12.037999
B    5.764629    4.603617   -11.438394
B    6.511205    4.548973   -10.775344
A    6.987514    3.704822   -10.529306
B    7.819385    4.211375   -10.756005
B    8.393817    4.441852    -9.970570
B    8.697227    3.490507   -10.024274
A    8.324763    2.816763    -9.386044
B    8.431279    2.794482    -8.391982
A    7.559624    2.419576    -8.076287
B    7.090471    3.228031    -7.720905
B    7.129588    4.108302    -8.193762
A    6.794454    3.948602    -9.122298
A    6.713603    3.124316    -9.682663
A    6.139630    2.723901   -10.396961
A    5.267798    3.014540   -10.791219
B    4.503855    2.386220   -10.938199
B    4.323784    2.515777    -9.963116
B    4.521673    3.479175   -10.143959
B    5.075225    4.062989   -10.737880
B    5.536667    4.879314   -10.390486
B    6.245523    4.684382    -9.712604
B    7.218278    4.576227    -9.917663
A    7.676200    3.711366    -9.711939
A    7.613535    2.857795   -10.229128
A    6.959172    2.114544   -10.089911
A    6.478405    1.761216   -10.892414
B    6.988109    1.005053   -11.302800
B    7.362086    0.196718   -10.848117
B    7.433998    0.151774    -9.851718
A    7.851979    0.980870    -9.480380
A    7.972920    1.876910    -9.907562
B    8.564523    2.432906   -10.491407
B    8.228152    3.223417   -11.003215
B    7.833432    2.633804   -11.707879
A    7.642734    1.991391   -10.965627
B    7.933594    1.092223   -10.638680
A    7.025067    1.046270   -10.223394
B    6.376887    0.309481   -10.415756
B    5.815884    0.956523   -10.932099
B    4.839685    0.739740   -10.938399
B    4.591087    1.453429   -10.283537
A    5.409067    1.958511   -10.558853
B    5.186689    1.664048   -11.488282
B    6.093856    1.482577   -11.867911
B    6.892296    2.082913   -11.822188
A    6.928889    2.779607   -11.105753
B    7.273290    3.603558   -11.555743
```